\documentclass[aps,pre,amsmath,amssymb,twocolumn,groupedaddress,showpacs]{revtex4}
\usepackage{graphicx}
\usepackage[utf8]{inputenc}

\usepackage{epsf}

\newcommand{\be}{\begin{equation}}
\newcommand{\ee}{\end{equation}}
\newcommand{\ba}{\begin{eqnarray}}
\newcommand{\ea}{\end{eqnarray}}
\newcommand{\baa}{\begin{eqnarray}}
\newcommand{\eaa}{\end{eqnarray}}
\newcommand{\ed}{\end{document}}

\renewcommand{\baselinestretch}{1.2}
\date{\today}
\begin{document}
\title{Transport in simple networks described by integrable discrete nonlinear Schr\"odinger equation}
\author{K. Nakamura$^{(1,4)}$, Z.A. Sobirov$^{(2)}$, D.U. Matrasulov$^{(2)}$, S. Sawada$^{(3)}$}
\affiliation{$^{(1)}$Faculty of Physics, National University of Uzbekistan, Vuzgorodok, Tashkent 100174,Uzbekistan\\
$^{(2)}$Turin Polytechnic University in Tashkent, 17 Niyazov Str., 100093, Tashkent, Uzbekistan\\
$^{(3)}$Department of Physics, Kwansei Gakuin University, Sanda 669-1337, Japan\\
$^{(4)}$Department of Applied Physics, Osaka City University, Osaka 558-8585, Japan}

\begin{abstract}
We elucidate the case in which the Ablowitz-Ladik (AL) type
discrete nonlinear Schr\"odinger equation (NLSE) on simple
networks (e.g., star graphs and tree graphs) becomes completely
integrable just as in the case of  a simple 1-dimensional (1-d)
discrete chain. 
The strength of cubic nonlinearity is
different from bond to bond, and networks are assumed to have  at
least two semi-infinite bonds with one of them working as an
incoming bond. 
The present work is a nontrivial extension of our preceding one (Sobirov {\it et al}, Phys. Rev. E {\bf 81}, 066602 (2010)) 
on the continuum NLSE to the discrete case.
We find: (1) the solution on each
bond is a part of the universal (bond-independent) AL soliton
solution on the 1-d discrete
chain, but is multiplied by the inverse of square root of
bond-dependent nonlinearity; (2)  nonlinearities at individual
bonds around each vertex must satisfy a sum rule; (3) under findings (1) and (2), there  exist  an infinite number of constants of
motion.  As a practical issue,  with use of AL soliton injected  
 through the incoming bond, we obtain transmission probabilities inversely proportional to the strength of nonlinearity on the outgoing bonds.\end{abstract}
\pacs{03.75.-b, 05.45.-a,05.60.Gg.}
\maketitle


\section{Introduction}
We shall investigate transport in networks with vertices and bonds which received
a growing  attention recently. The networks of practical importance are those 
of nonlinear waveguides and and optical fibers \cite{kiv},
double helix of DNA \cite{yak}, 
Josephson junction arrays with Bose-
Einstein Condensates \cite{cat},
vein networks in leaves
\cite{cor,kat}, etc.

Major theoretical concern so far, however, is limited to solving stationary states of the linear
Schr\"odinger  equation, and to obtaining the  energy spectra in closed
networks and transmission probabilities for open networks with
semi-infinite leads \cite{exn0,exn,kott,kura,kuch,exner}.
Only a few studies treats  the nonlinear
Schr\"odinger  equation on simple networks, which are still limited to
the analysis of its stationary state\cite{gnut,adam}.

With  introduction of the nonlinearity to the time-dependent
Schr\"odinger equation, the network becomes to provide a nice
playground where one can see interesting soliton propagations and
nonlinear dynamics through the network
\cite{fla,bur,pik,fla2}, namely through an assembly of continuum line
segments connected at vertices. Although there exist important analytical studies 
on the semi-infinite and finite chains \cite{ablo,foka1,ramos,foka2}, 
we find little exact
analytical treatment of soliton propagation through networks
 within a framework of nonlinear Schr\"odinger
equation(NLSE) \cite{sulem,abl}. The subject is difficult due to the
presence of vertices where the underlying chain should bifurcate
or multi-furcate in general.

Recently, with a suitable boundary condition at each vertex we
developed an exact analytical treatment of soliton propagation
through networks within a framework of NLSE \cite{sob}. Under an
appropriate   relationship among values of nonlinearity at
individual bonds, we found nonlinear dynamics of solitons with no
reflection at the vertex. We also showed that an infinite number
of constants of motion are available for NLSE on networks, namely
the mapping of Zakharov-Shabat (ZS)'s scheme\cite{zakh} to networks was achieved.

The extension of the scenario to  the discrete NLSE (DNLSE) is far
from being obvious. The standard DNLSE is not integrable and the
integrable variant of the continuum nonlinear Schr\"odinger
equation is the one proposed by Ablowitz and Ladik (AL)\cite{abl2,abl3,abl,bio}. 
AL equation is the appropriate choice for the zeroth order approximation in studying the soliton dynamics perturbatively in physically motivated models, such as an array of coupled optical waveguides \cite{ace} and proton dynamics in hydrogen-bonded chains \cite{yomo,malo}.
The dynamics of intrinsic localized modes in nonlinear lattices can be
approximately described by AL equation \cite{take}. Exciton systems
with exchange and dipole-dipole interactions also reduce to AL
equation in some limiting cases\cite{kono}. The AL chain  is integrable by means of the inverse
scattering transform, and, together with the  Toda lattice \cite{toda}, constitutes a paradigm of the completely integrable lattice systems.

 AL equation for a field variable $\psi$ on one-dimensional
(1-d) chain is given by
\begin{align}\label{al-0}
i\dot\psi_{n}+\left(\psi_{n+1}+\psi_{n-1}\right)\left(1+\gamma|\psi_{n}|^2\right)=0,
\end{align}
where $\gamma$ is the strength of nonlinear inter-site interaction
and $n$ denotes each lattice site on the chain. This equation can be
obtained from the canonical equation of motion with use of the
non-standard Poisson brackets. Equation (\ref{al-0}) has an
infinite number of independent constants of motion and is
completely integrable \cite{abl2,abl3}.

However, there is an ambiguity in  generalizing the AL model to
networks: how can we define the inter-site interaction at each
vertex in order to see the infinite-number of constants of motion
in networks? To keep the integrability of AL equation, should any 
rule hold for the strength of nonlinearity on bonds
joining at each vertex? We shall resolve these questions in this
paper and show how solitons of AL equation on networks will be mapped to that of AL equation on a 1-d chain.
Once this mapping will be found, the integrability properties like the inverse scattering transform, B\"acklund transformation, etc, are automatically 
guaranteed, and will not be addressed in this paper.

Below we shall show the completely integrable case of the AL
equation on networks with strength of nonlinearity different from bond to bond. As a relevant issue,  with use of reflectionless propagation of AL soliton through networks, we shall evaluate the transmission probabilities on the outgoing bonds.
In Section II, using a primary star
graph (PSG) and defining a suitable equation of motion at the vertex, we shall address the norm and energy conservations.
In Section III,  we shall show a basic idea of the soliton propagation
along the branched chain, finding the connection formula at the vertex and the sum rule among the strengths of
nonlinearity on the bonds, which guarantee the
infinite number of constants of motions and complete integrability
of the system under consideration.
In Section IV, the cases
of generalized star graphs and tree graphs are investigated.  Section V is devoted to
the investigation of an injection of  AL soliton which
bifurcates at the vertex and is decomposed into a pair of solitons
with each propagating along the outgoing bonds, and we shall evaluate the
transmission probabilities on the outgoing bonds. Summary and discussions are given in
Section VI.


\section{Norm and energy conservations on primary star graph}
\subsection{Ablowitz-Ladik(AL) equation on networks}
\begin{figure}[htb]
\centerline{\includegraphics[width=\columnwidth]{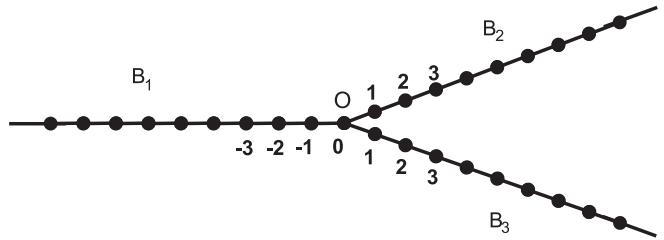}}
\caption{Primary star graph.  3 semi-infinite chains $B_1, B_2$
and $B_3$ connected at a vertex $O$.} \label{star}
\end{figure}

Let us consider an elementary branched chain (see Fig.\ref{star}), namely, a primary star graph (PSG) consisting of three semi-infinite bonds connected at the vertex $O$.

We denote individual lattice sites as $(k,n)$, where $k=1, 2$, and
$3$ mean the bond's number and $n$ corresponds to a lattice site on
each bond. For the first bond $k=1$, $n$ is numbered as $n\in
B_1=\{0,-1,-2,\cdots\}$, where $(1,0)$ means the branching point,
i.e., the vertex. For the second ($k=2$) and third ($k=3$) bonds,
$n$ varies as  $n\in B_k =\{1,2,3,\cdots\}$. $(2,1)$ and $(3,1)$
stands for the points nearest to the vertex.

Discrete nonlinear Schro\"odinger equation (DNLSE) {\it \'{a} la}
Ablowitz-Ladik (AL) is defined on each bond except for the vicinity of the vertex
as
\begin{align}\label{eq1}
i\dot\psi_{k,n}+\left(\psi_{k,n+1}+\psi_{k,n-1}\right)\left(1+\gamma_k|\psi_{k,n}|^2\right)=0,
\end{align}
where $(k,n)\not\in\{(1,0), (2,1), (3,1)\}$. It should be noted
that $\gamma_k$ may be different among bonds. There is an
ambiguity about the interaction around the vertex, which is
resolved as follows: Let's first introduce Hamiltonian for PSG as
\begin{align}\label{Net-Haml}
H=-\sum_{n=0}^{-\infty}(\psi_{1,n}^*\psi_{1,n+1} + c.c.)-\sum_{k=2}^{3}\sum_{n=1}^{+\infty}(\psi_{k,n}^*\psi_{k,n+1} + c.c.),
\end{align}
where at the virtual site $(1,1)$ we assume
$\psi_{1,1}=s_2\psi_{2,1}+s_3\psi_{3,1}$ with appropriate
coefficients $s_2$ and $s_3$. Then Eq.(\ref{eq1}) can be obtained
by the equation of motion
\begin{align}\label{Net-al}
i\dot\psi_{k,n}=\{H,\psi_{k,n}\}
\end{align}
at $(k,n)\not\in\{(1,0), (2,1), (3,1)\}$, with use of  non-standard Poisson brackets
\begin{align}\label{Net-PB}
\{\psi_{k,m},\psi^*_{k^\prime,n}\}=i(1+\gamma|\psi_{k,m}|^2)\delta_{kk^{\prime}}\delta_{mn},\nonumber\\
\{\psi_{k,m},\psi_{k^\prime,n}\}=\{\psi^*_{k,m},\psi^*_{k^\prime,n}\}=0.
\end{align}
On the same footing as above, the equation of motions in
Eq.(\ref{Net-al})  at (1,0), (2,1) and (3,1) are given,
respectively, as
\begin{align}\label{v10}
i\dot\psi_{1,0}+\left(\psi_{1,-1}+s_2\psi_{2,1}+s_3\psi_{3,1}\right)\left(1+\gamma_1|\psi_{1,0}|^2\right)=0,
\end{align}
\begin{align}\label{vk1}
i\dot\psi_{k,1}+\left(s_k\psi_{1,0}+\psi_{k,2}\right)\left(1+\gamma_k|\psi_{k,1}|^2\right)=0,\
\ k=2,3.
\end{align}

The solution is assumed to satisfy the following conditions at
infinity: $\psi_{1,n}\to 0$ at $n\to -\infty$ and  $\psi_{k,n}\to
0$ at $n\to +\infty$ for $k=2$ and 3.


\subsection{Norm and energy conservations}
It is known that the  norm conservation is  one of the most
important physical conditions in conservative systems. Since
Eqs.(\ref{eq1}),(\ref{v10}) and (\ref{vk1})  are available from
Hamilton's equation of motion with  non-standard Poisson brackets,
the norm and energy conservations seem obvious. Below, however, we observe them
explicitly. Extending the definition in the case of 1-d chain
\cite{abl}, the norm for PSG is given as
\begin{align}\label{Net-norm}
N=\|\psi\|^2=\sum_{k=1}^3\frac{1}{\gamma_k}\sum_{n\in
B_k}\ln\left(1+\gamma_k|\psi_{k,n}|^2\right).
\end{align}
Its time derivative is given by
\begin{align}\label{dotN}
\frac{d}{dt}N=\sum_{k=1}^3\sum_{n\in
B_k}A_{k,n}
\end{align}
with
\begin{align}\label{Akn}
A_{k,n}=\frac{1}{1+\gamma_k|\psi_{k,n}|^2}\left(\psi_{k,n}^*\dot\psi_{k,n}+\dot\psi_{k,n}^*\psi_{k,n}\right).
\end{align}
For $(k,n)\not\in \{(1,0), (2,1), (3,1)\}$ with use of Eq.
(\ref{eq1}) we have
\begin{align}\label{Akn1}
A_{k,n}&=\frac{1}{i}\left(\psi_{k,n}\psi_{k,n+1}^*-\psi_{k,n}^*\psi_{k,n+1}\right)\nonumber\\
&-\frac{1}{i}\left(\psi_{k,n-1}\psi_{k,n}^*-\psi_{k,n-1}^*\psi_{k,n}\right)\nonumber\\
&\equiv j_{k,n}-j_{k,n-1},
\end{align}
where
\begin{align}\label{current}
j_{k,n}\equiv \frac{1}{i}\left(\psi_{k,n}\psi_{k,n+1}^*-\psi_{k,n}^*\psi_{k,n+1}\right)
\end{align}
implies a local current. Firstly one observes
\begin{align}
\sum\limits_k{\sum\limits_n}^\prime
A_{k,n}=j_{1,0}-j_{2,1}-j_{3,1}, \label{sum-up-Akn}
\end{align}
where $\sum\limits_k{\sum\limits_n}^\prime$ means the summation over
all sites on PSG except for the points $(1,0), (2,1),
(3,1)$.

Then, for $(k,n)= (1,0), (2,1), (3,1)$, with use of Eqs.
(\ref{v10}) and (\ref{vk1}) we obtain
\begin{align}\label{A10}
A_{1,0}&=s_2\frac{1}{i}\left(\psi_{1,0}\psi_{2,1}^*-\psi_{1,0}^*\psi_{2,1}\right)\nonumber\\
&+s_3\frac{1}{i}\left(\psi_{1,0}\psi_{3,1}^*-\psi_{1,0}^*\psi_{3,1}\right)
-j_{1,0}
\end{align}
and
\begin{align}\label{Ak1}
A_{k,1}=j_{k,1} - s_k\frac{1}{i}\left(\psi_{1,0}\psi_{k,1}^*-\psi_{1,0}^*\psi_{k,1}\right)
\end{align}
for $k=2,3$. Substituting Eqs.(\ref{sum-up-Akn}), (\ref{A10}) and (\ref{Ak1}) into Eq.(\ref{dotN}), we can see
$\frac{d}{dt}N=0$, i.e., the norm conservation.
Therefore, for any choice of values $s_2$ and $s_3$ the norm
conservation turns out to hold well.


On the other hand, the energy for PSG is expressed in a symmetrical form as
\begin{align}\label{Energy1}
E=-2 { \rm Re}&
\left[\sum_{n=-1}^{-\infty}\psi_{1,n}^*\psi_{1,n+1}+\sum_{k=2}^{3}\sum_{n=1}^{+\infty}\psi_{k,n}^*\psi_{k,n+1}\right.\nonumber\\
+&\left. \psi_{1,0}^*(s_2\psi_{2,1}+s_3\psi_{3,1})\right].
\end{align}

To show that the energy is conservative, 
we see its time derivative
\begin{align}\label{dotE}
\frac{d}{dt}E=&-2 { \rm Re}
\sum_{n=-1}^{-\infty}\left(\psi_{1,n}^*\dot\psi_{1,n+1}+\dot\psi_{1,n}^*\psi_{1,n+1}\right)\nonumber\\
-& 2{\rm Re}
\sum_{k=2}^{3}\sum_{n=1}^{+\infty}\left(\psi_{k,n}^*\dot\psi_{k,n+1}+\dot\psi_{k,n}^*\psi_{k,n+1}\right)-\nonumber\\
-&2{\rm Re}\left[
\psi_{1,0}^*\left(s_2\dot\psi_{2,1}+s_3\dot\psi_{3,1}\right)+\dot\psi_{1,0}^*\left(s_2\psi_{2,1}+s_3\psi_{3,1}\right)
\right].
\end{align}

With use of Eq. (\ref{eq1}) we have
\begin{align}\label{Esum1}
-&\sum_{n=-1}^{-\infty}\left(\psi_{1,n}^*\dot\psi_{1,n+1}+\dot\psi_{1,n}^*\psi_{1,n+1}\right)\nonumber\\
=&
\frac{1}{i}\sum_{n=-1}^{-\infty}\left[|\psi_{1,n-1}|^2-|\psi_{1,n+1}|^2\right]\left(1+\gamma_1|\psi_{1,n}|^2\right)\nonumber\\
-&\psi_{1,-1}^*\dot\psi_{1,0},
\end{align}
and
\begin{align}\label{Esum23}
-&\sum_{n=1}^{\infty}\left(\psi_{k,n}^*\dot\psi_{k,n+1}+\dot\psi_{k,n}^*\psi_{k,n+1}\right)\nonumber\\
=&
\frac{1}{i}\sum_{n=2}^{\infty}\left[|\psi_{k,n-1}|^2-|\psi_{k,n+1}|^2\right]\left(1+\gamma_1|\psi_{k,n}|^2\right)-\dot\psi_{k,1}^*\psi_{k,2}.
\end{align}
The first terms in the final expressions in Eqs.(\ref{Esum1}) and (\ref{Esum23}) are obviously pure-imaginary. 
Substituting Eqs. (\ref{Esum1}) and (\ref{Esum23}) into Eq.(\ref{dotE}) and
using Eqs.(\ref{v10}) and (\ref{vk1}), we find:
\begin{align}\label{dotE2}
&\frac{d}{dt}E=-2{\rm
Re}\left[\psi_{1,0}^*\left(s_2\dot\psi_{2,1}+s_3\dot\psi_{3,1}\right)+\right.\nonumber\\
\dot\psi_{1,0}^* &
\left(s_2\psi_{2,1}+s_3\psi_{3,1}\right)+\psi_{1,-1}^*\dot\psi_{1,0}
+\nonumber\\
&+\left.
\dot\psi_{2,1}^*\psi_{2,2}+\dot\psi_{3,1}^*\psi_{3,2}\right]\nonumber\\
=2{\rm Re}&
\left[\frac{1}{i}(1+\gamma_1|\psi_{1,0}|^2)\left(|\psi_{1,-1}|^2-|s_2\psi_{2,1}+s_3\psi_{3,1}|^2\right)\right.\nonumber\\
+
\frac{1}{i}&\left.\sum_{k=1}^3(1+\gamma_k|\psi_{k,1}|^2)\left(s_k^2|\psi_{1,0}|^2-|\psi_{k,2}|^2\right)\right]=
0.
\end{align}
The last equality comes from the pure-imaginary nature of the
expression in $[\cdots]$. Equation (\ref{dotE2}) is nothing but the energy conservation.

Thus we have proved that the norm
and energy are conserved for any choice of values $s_2$
and $s_3$. In general, however, other conservation rules do not hold.
In the next sections we shall reveal a special case with
appropriate choice of $s_2$ and $s_3$ which guarantees an infinite
number of conservation laws.


\section{Completely Integrable Case}
\subsection{Dynamics near branching point and sum rule}

\begin{figure}[htb]
\centerline{\includegraphics[width=\columnwidth]{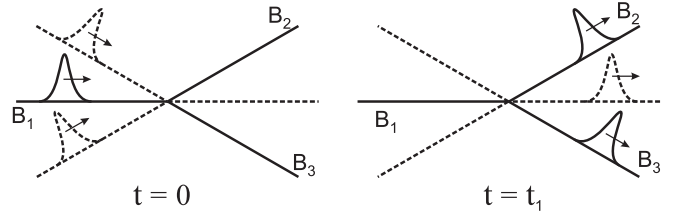}}
\caption{Real bonds and real solitons (solid lines) and ghost bonds and ghost solitons  (broken lines).}
\label{ghost}
\end{figure}

  Among many possible choices of  $s_2$ and $s_3$, there is one
special case in which an infinite number of constants of motion
can be found and DNLSE on PSG becomes completely integrable. To investigate this case,
we shall first add to each bond $B_k$ ($k=1,2,3$) a ghost-bond counterpart $B_k^\prime$ so that  $B_k +  B_k^\prime$ constitutes an ideal 1-d chain (see Fig. \ref{ghost}). 
Then we suppose that the soliton solution of AL equation on PSG is given by
\begin{align}
\psi_{k,n}(t)=\frac{1}{\sqrt{\gamma_k}}q_{k,n}(t), \ \ k=1,2,3
\label{GSol-1}
\end{align}
where $q_{k,n}(t)$ are soliton solutions of DNLSE with unit nonlinearity on the ideal 1-d chain (\cite{abl,abl2,abl3}):
\begin{align}
i\dot q_n +(q_{n+1}+q_{n-1})(1+|q_n|^2)=0,  \label{AL on line}
\end{align}
with $n$ being integers in $(-\infty , +\infty)$. The solutions of Eq.(\ref{AL on line}) may be different among three fictitious chains
 $B_k + B_k^\prime$  ($k=1,2,3$).

Comparing Eqs. (\ref{v10}), (\ref{vk1}) and (\ref{AL on line}),
one can find at the vertex the following two equalities:
\begin{align}
\frac{1}{\sqrt{\gamma_1}}q_{1,1}(t)=\frac{s_2}{\sqrt{\gamma_2}}q_{2,1}(t)
+ \frac{s_3}{\sqrt{\gamma_3}}q_{3,1}(t), \label{Comp-nv10}
\end{align}
\begin{align}\frac{1}{\sqrt{\gamma_k}}q_{k,0}(t)=\frac{s_k}{\sqrt{\gamma_1}}q_{1,0}(t),
\ \ k=2,3.\label{Comp-nvk1}
\end{align}

Noting the spatio-temporal behavior of soliton solutions and to guarantee the equality in Eq.(\ref{Comp-nvk1}), 
$q_{k,n}(t)=s_k\sqrt\frac{\gamma_k}{\gamma_1}q_{1,n}(t)$ with $k=2,3$ should be satisfied for any time $t$ and for any integer $n$, from which
we obtain
\begin{align}
s_k\sqrt\frac{\gamma_k}{\gamma_1}=1\ \ \mbox{ or  }
s_k=\sqrt\frac{\gamma_1}{\gamma_k}\ \ (k=2,3) \label{S-k}
\end{align}
and
\begin{align}
q_{k,n}(t)\equiv q_{n}(t) \label{q-kn},
\end{align}
namely, the solution $q_{k,n}(t)$ should be bond-independent.
With use of Eqs.(\ref{S-k}) and (\ref{q-kn}) in Eq. (\ref{Comp-nv10}) we
have the sum rule among nonlinearity coefficients $\gamma_1$, $\gamma_2$ and $\gamma_3$:
\begin{align}
\frac{1}{\gamma_1}=\frac{1}{\gamma_2}+\frac{1}{\gamma_3}
\label{sumrule-g}
\end{align}

Equations (\ref{S-k}), (\ref{q-kn}) and (\ref{sumrule-g}) are the necessary and sufficient conditions to see Eqs.(\ref{Comp-nv10}) and (\ref{Comp-nvk1}). Thus, under the sum rule for nonlinearity coefficients in Eq.(\ref{sumrule-g}),
the solution on PSG is given by a common (bond-independent) soliton solution of 
Eq.(\ref{AL on line}) multiplied by square root of the inverse
nonlinearity coefficient. For example, the soliton incoming through the bond $B_1$ is expected to
smoothly bifurcate at the vertex and propagate through the bonds $B_2$ and $B_3$, as we shall see in Fig. \ref{numerical}.
In the case that $\gamma_1$, $\gamma_2$ and $\gamma_3$ break the sum rule,  we
shall see a completely different nonlinear dynamics of solitons
such as their reflection and emergence of radiation  at the vertex, as will be shown in Fig.\ref{numerical-ex}.  The initial value
problem for such a case is outside the scope of the present work.

We also note that the parameters $s_2$ and $s_3$  would correspond to
$\frac{\alpha_2}{\alpha_1}$ and $\frac{\alpha_3}{\alpha_1}$
, respectively, in the preceding work \cite{sob}, although the derivations of the connection formula at the vertex
are quite different between the continuum and discrete systems. In fact, $s_2$ and $s_3$ are introduced to define the inter-site interaction at the vertex and are not obtained  from the norm and energy conservations, in contrast to the case of networks consisting of continuum segments\cite{sob}.


\subsection{An Infinite number of constants of motion }

It is well known that Ablowitz-Ladik (AL) equation on the 1-d
chain has an infinite number of constants of motion. Now we shall
proceed to obtain an infinite number of constants of motion for
general solutions of AL equation on PSG. First of all, it should
be noted that the solution on PSG can now be
written as
\begin{align}\label{genSol}
\psi_{k}(t)=\frac{1}{\sqrt {\gamma_k}} \{q_{n}(t)|n\in B_k\}, \qquad
k=1,2,3,
\end{align}
where  $q(t)$ stands for a general solution of
AL equation (\ref{AL on line}) and is restricted to each bonds $B_k$ ($k=1,2,3$).

While we already proved the conservation of
energy,  we can generalize it to the general case:
Without taking the complex conjugate,  Eq. (\ref{Net-Haml}) can be
explicitly written as
\begin{align}\label{gl-Energy1}
Z=-\sum_{n=-1}^{-\infty}\psi_{1,n}^*\psi_{1,n+1}&-\sum_{k=2}^{3}\sum_{n=1}^{+\infty}\psi_{k,n}^*\psi_{k,n+1} \nonumber\\
&-\psi_{1,0}^*(s_2\psi_{2,1}+s_3\psi_{3,1}).
\end{align}
Substituting Eq.(\ref{genSol}) into Eq.(\ref{gl-Energy1}), $Z$  is rewritten as
\begin{align}\label{gl-Energy2}
Z=&-\frac{1}{\gamma_1}\sum_{n=0}^{-\infty}q_{n}^*q_{n+1}-\sum_{k=2}^{3}\frac{1}{\gamma_k}\sum_{n=1}^{+\infty}q_{n}^*q_{n+1} \nonumber\\
&+ \frac{1}{\gamma_1}q_{0}^*q_{1}  -\sum_{k=2}^{3}\frac{s_k}{\sqrt{\gamma_1\gamma_k}}q_{0}^*q_{1}.
\end{align}
Using the value $s_k$ in Eq.(\ref{S-k}) and the sum rule in
Eq.(\ref{sumrule-g}), Eq.(\ref{gl-Energy2}) reduces to the
constant for the ideal 1-d chain \cite{abl2, abl3}:
\begin{align}\label{gl-Energy3}
Z=-\frac{1}{\gamma_1}\sum_{-\infty}^{+\infty}q_{n}^*q_{n+1}.
\end{align}
Therefore $Z$ in Eq.(\ref{gl-Energy1}) is a constant of motion, and its real and imaginary parts imply the energy and current, respectively.

For other higher-order conservation rules, we can write them as
\begin{align}\label{ConsLaws}
\frac{1}{\gamma_1}C_m=&\frac{1}{\gamma_1}\sum_{n=0}^{-\infty}f_m^{(n)}(\{q_n|n\in B_{1}\})\nonumber\\
&+\sum_{k=2}^3\frac{1}{\gamma_k}\sum_{n=1}^{+\infty}f_m^{(n)}(\{q_n|n\in B_{k}\}),
\end{align}
with $f_m$ defined as expansion coefficients of the
expression (see Ablowitz \& Ladik \cite{abl3})
\begin{align}\label{ConsLaws2}
\log (g_n^{(0)}+g_n^{(1)}z^2+g_n^{(2)}z^4+\cdots
)=f_1^{(n)}z^2+f_2^{(n)}z^4+\cdots \ ,
\end{align}
where $(g_{n}^{(m)})$ are given by
\begin{align}\label{CL3}
&g_n^{(0)}=1, \ \ g_n^{(1)}=R_{n-1}Q_{n-2},\nonumber\\
&g_{n}^{(m)}=\frac{R_{n-1}}{R_{n-2}}g_{n-1}^{(m-1)}-\sum_{l=1}^{m-1}g_{n-1}^{(m-l)}g_{n}^{(l)},\
\ m=2,3,4,\cdots,
\end{align}
\begin{align}\label{CL4}
R_n=q_{n+2}^*,\ \ Q_n=-q_{n+2}.
\end{align}
The relations (\ref{CL3}) and (\ref{CL4}) are obtained by solving Eq. (4.15) in \cite{abl3}, i.e.,
\begin{align}\label{g-rec}
g_{n+1}(g_{n+2}-1)-z^2\frac{R_{n+1}}{R_n}(g_{n+1}-1)=z^2R_{n+1}Q_n,
\end{align}
recursively with use of the expansion
\begin{align}\label{g-n}
g_n=g_n^{(0)}+g_n^{(1)}z^2+g_n^{(2)}z^4+\cdots .
\end{align}

The right-hand side of Eq. (\ref{ConsLaws})
includes some  undefined field variables in the ghost bond regions which  must be defined as
\begin{align}\label{undef}
&\psi_{1,n}=\sqrt {\frac{\gamma_1}{\gamma_2}}\psi_{2,n}+\sqrt {\frac{\gamma_1}{\gamma_3}}\psi_{3,n} \qquad \text{  with  } n\geq 1,\nonumber\\
&\psi_{k,n}=\sqrt {\frac{\gamma_1}{\gamma_k}}\psi_{k,n}, \quad k=2,3 \qquad \text{ with } n\leq 0.
\end{align}
The conservation laws in Eq. (\ref{ConsLaws}) follows from the
nature of solutions (\ref{genSol}) and the sum rule for
nonlinearity coefficients (\ref{sumrule-g}).

For $m=1$  we obtain current and energy conservation laws. At
$m\geq 2$ we obtain higher order conservation laws.
Some of higher-order constants of motion are as follows:
\begin{align}\label{cm2}
\frac{1}{\gamma_1}C_2=&-\sum_{k=1}^{3}\sum_{n\in B_k}\left(
\psi_{k,n+1}^*\psi_{k,n-1}(1+\gamma_k|\psi_{k,n}|^2)\right.\nonumber\\
&\left.+\frac{\gamma_k}{2}\psi_{k,n}^2(\psi_{k,n+1}^*)^2\right),
\end{align}

\begin{align}\label{cm3}
\frac{1}{\gamma_1}C_3=&-\sum_{k=1}^{3}\sum_{n\in B_k}\left[(
\psi_{k,n+2}^*\psi_{k,n-1}(1+\gamma_k|\psi_{k,n+1}|^2)+\right.\nonumber\\
&+\gamma_k\psi_{k,n}^*\psi_{k,n+1}^*\psi_{k,n-1}^2
\nonumber\\ &+(\psi_{k,n+1}^*)^2\psi_{k,n}\psi_{k,n-1})(1+\gamma_k|\psi_{k,n}|^2)\nonumber\\
&+\frac{\gamma_k^2}{3}\left. \psi_{k,n+1}^*\psi_{k,n}\right],
\end{align}
where field variables at lattice sites of the ghost bonds are defined in Eq. (\ref{undef}).


\section{Generalized star and tree graphs}

  Now we proceed to explore soliton solutions of DNLSE on
other types of graphs and explore the sum rule and conservation
rules for solitons to propagate through these graphs.

The above treatment on PSG is also true for more  general star graphs
consisting of $N$ semi-infinite bonds connected at a single
vertex. In such cases, the initial soliton at an incoming bond $B_1$ splits into
$N-1$ solitons in the remaining bonds, and the extended version of
Eq. (\ref{sumrule-g}) is
\begin{align}\label{beta-g}
\frac{1}{\gamma_1}=\sum_{j=2}^{N}\frac{1}{\gamma_j}.
\end{align}
The solution is given by the equations
\begin{align}
\psi_{k,n}(t)=\frac{1}{\sqrt{\gamma_k}}q_n(t),
\label{sol-gen-star}
\end{align}
where $n=0,-1,-2,\cdots$ for the first bond ($k=1$) and $n=1,2,3,\cdots$
for other bonds ($2\leq k\leq N$). $q_n(t)$ is a soliton solution
of Eq. (\ref{AL on line}). Conservation laws for this graph
can be obtained analogously as in the case of PSG.
\begin{figure}[htb]
\centerline{\includegraphics[width=\columnwidth]{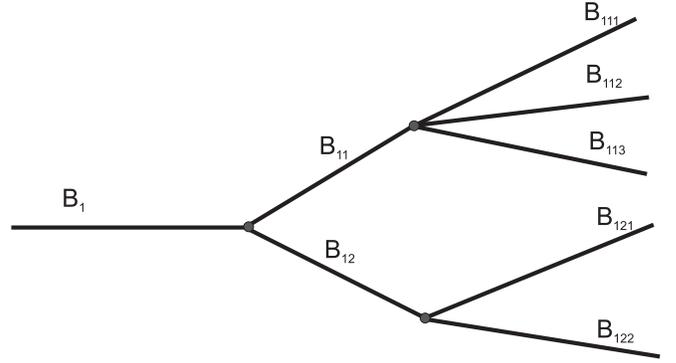}}
\caption{Tree graph. $B_1\sim (-\infty, 0), B_{11}, B_{12} \sim
(0,L)$, and $B_{1ij} \sim (0,+\infty)$ with $i,j=1,2, \cdots$.}
\label{tree-0}
\end{figure}

Another example of the graph for which the soliton solution of
DNLSE can be obtained analytically is a tree graph  in Fig.
\ref{tree-0}.
Now we shall provide a soliton solution in this case. We denote
bonds of graph as $B_{\Lambda}=B_{1ij \cdots m}$ and  number the lattice sites on this bonds as $1,2,3,\cdots,N_{\Lambda}$.
On each branching point we assume the following conditions are
hold
\begin{align}\label{sumrule}
\frac{1}{\gamma_{\Lambda}}=\sum_m \frac{1}{\gamma_{\Lambda m}}.
\end{align}
The solution is given by
\begin{align}\label{treesoliton}
\psi_{\Lambda,
n}(t)=\frac{1}{\sqrt{\gamma_k}}q_{n+s_{\Lambda}}(t), \qquad n\in
B_{\Lambda}.
\end{align}
Here $s_{\Lambda}$ number of lattice sites that soliton pass through from
$B_1$ to $B_{\Lambda}$. For the tree graph it is defined as
\begin{align}\label{sdef}
&s_1=s_{1i}=n_0,\ s_{1ij}=n_0+N_{1i}, \nonumber\\
&s_{\Lambda}\equiv s_{1ij \cdots lm}=n_0+N_{1i}+\cdots+N_{1ij \cdots l}.
\end{align}

Below, applying the induction method we give a proof of
conservation laws for soliton solutions of AL on tree graph. Let us
denote the tree graph as $G$ and assume the  conservation laws to
hold in $G$: $\sum_{B_{\Lambda}\in G} \sum_{n\in
B_{\Lambda}}f_n^{(k)}(q_{n+s_{\Lambda}}(t))=const$. Then we
construct an enlarged tree graph in the following way: First we
choose the arbitrary point $N_{\Phi}$ in the one of right-most
semi-infinite chain $B_{\Phi}$ as a new branching point. Cut of
semi-infinite part of this bond at the point $N_{\Phi}$ and attach
$M$ semi-infinite bonds to this point. Namely the bond $B_{\Phi}$
is now replaced by a finite bond $\tilde B_{\Phi}$ connected with
$M$ semi-infinite bonds $B_{\Phi m}=\{1,2,\cdots,N_{\Phi m}\}$, with
$m=1,2,\cdots,M$. For the enlarged tree graph, constants of motion are
given by
\begin{align}\label{ind}
&\sum_{B_{\Lambda}\in G-B_{\Phi}}\gamma_{\Lambda}^{-1} \sum_{n\in
B_{\Lambda}}f_n^{(k)}(q_{n+s_{\Lambda}}(t))+
\gamma_{\Phi}^{-1}\sum_{n\in
\tilde B_{\Phi}}f_n^{(k)}(q_{n+s_{\Phi}}(t))\nonumber\\
&+ \sum_{m=1}^M\gamma_{\Phi m}^{-1}\sum_{n\in
B_{\Phi m}}f_n^{(k)}(q_{n+s_{\Phi}+N_{\Phi}}(t))\nonumber\\
&=\sum_{B_{\Lambda}\in G-B_{\Phi}}\gamma_{\Lambda}^{-1} \sum_{n\in
B_{\Lambda}}f_n^{(k)}(q_{n+s_{\Lambda}}(t))+
\gamma_{\Phi}^{-1}\sum_{n=1}^{N_{\Phi}}f_n^{(k)}(q_{n+s_{\Phi}}(t))\nonumber\\
&+\sum_{m=1}^M\gamma_{\Phi m}^{-1}\sum_{n=1+N_{\Phi
m}}^{+\infty}f_n^{(k)}(q_{n+s_{\Phi}+N_{\Phi}}(t))\nonumber\\
&=-\left(\gamma_{\Phi}^{-1}-\sum_{m=1}^M\gamma_{\Phi
m}^{-1}\right)\sum_{n=1+N_{\Phi
m}}^{+\infty}f_n^{(k)}(q_{n+s_{\Phi}+N_{\Phi}}(t))+const.
\end{align}

It is clear that the final expression becomes constant under the
sum rule (\ref{sumrule}). Thus, starting from PSG in Fig. \ref{star} and repeating the above procedure, we can get the conservation rule for all tree graphs.


\section{Transmission probabilities against injection of a single soliton}

A relevant issue of the above discoveries is the transmission probability against injection of a single soliton.  
Here we calculate transmission probabilities for a single soliton
which is incoming through a semi-infinite bond $B_1$ and outgoing
through the other semi-infinite bonds $\{B_l|l \ne 1\}$.

A single (bright) soliton on a graph, which takes the general form
as in Eqs. (\ref{genSol}), (\ref{sol-gen-star}) and (\ref{treesoliton}), is described with use of AL soliton with
$\gamma =1$ \cite{abl2}: $\psi_{l,n}(t)$ lying on individual bonds $B_l$ is
given by
\begin{align}\label{Sol}
\psi_{l,n}(t)=&\gamma_l^{-1/2}\sinh\beta {\rm sech}
[\beta(n-n_{0}-vt)] \nonumber\\
&\times e^{-i(\omega t+\alpha
n+\phi_{0})}, \nonumber\\
& \qquad n\in B_l, \ l=1,2,3, \cdots, N,
\end{align}
where $ \omega =-2{\rm cosh}\beta\cos\alpha$, $v=-(2/\beta){\rm
sinh}\beta\sin\alpha$, $-\pi\leq\alpha\leq\pi$, $0<\beta<\infty$, $0\leq\phi_{0}<2\pi$ and $n_{0}$ are bond-independent parameters
characterizing frequency, velocity, wave number, inverse width of
the soliton, initial phase and initial  center of mass, respectively. 
Equation (\ref{Sol}) indicates that a narrow soliton travels faster than wider ones with the same $\alpha$.

It should be noted that parameter values are common to each bond, except for $\{\gamma_l\}$. Choosing the simplest network PSG in Fig.\ref{star}, we shall give
conservative quantities for the solution in Eq.  (\ref{Sol}) under the sum rule in Eq.(\ref{sumrule-g}). First of
all, the norm in Eq.(\ref{Net-norm}) turned out to be reduced to the one for the 1-d chain with the nonlinearity constant $\gamma_1$ and thereby is given by 
\begin{align}\label{norm-sol}
N=2\beta/\gamma_1.
\end{align}
Equation (\ref{norm-sol}) indicates that a narrow soliton has  a larger norm than wider ones.
As for the energy $(E)$ and current $(J)$, it is
convenient to evaluate the combined quantity $Z$ in Eq. (\ref{gl-Energy1}) with use of $s_2$ and $s_3$ given by Eq. (\ref{S-k}).
In fact we have
\begin{align}\label{EnCur-1}
E=-2 {\rm Re} ( Z), \qquad J=2 {\rm Im} ( Z).
\end{align}

Substituting Eq. (\ref{Sol}) into Eq. (\ref{gl-Energy1}) and using the sum rule in Eq. (\ref{sumrule-g}), 
one obtains
\begin{align}\label{EnCur3}
Z=\frac{2}{\gamma_1}e^{-i\alpha}\sinh\beta
\end{align}
and
\begin{align}\label{En3}
E=-\frac{4}{\gamma_1}\cos{\alpha}\sinh\beta,\qquad
J=-\frac{4}{\gamma_1}\sin{\alpha}\sinh\beta.
\end{align}

As  is seen from Eq. (\ref{Sol}), the center of mass of the
soliton (CMS) on each bond $B_l$ is located at $n =
n_0$  at $t=0$. However, lattice points on the individual
semi-infinite bonds are defined on the limited interval. In particular, 
on outgoing bonds $\{B_l|l \ne 1\}$, their lattice points
$n$ are defined in the interval $(1, +\infty)$. If
$n_0 < 0$, therefore, CMS on $\{B_l|l \ne 1\}$ is initially located
outside of the real bonds. In such cases we call the soliton as a
"ghost soliton". When CMS belongs to a real bond we use a term "real
soliton". In Fig. \ref{ghost} which corresponds to PSG in Fig.\ref{star}, ghost solitons are plotted by broken
curve while real ones by solid line. The soliton
dynamics here is governed by a single characteristic time $\tau
\equiv \frac{-n_0}{v}$. While for $0\le t \le \tau$ the soliton at
$B_1$  is a real one and those at $B_2$ and $B_3$ are ghosts, for
$\tau \le t$ the soliton at $B_1$ is a ghost and those at $B_2$
and $B_3$ are real. 
At $t=0$ with $-n_0 \gg 1$, the soliton lying on the bond $B_1$
is exclusively responsible for the norm $N$. On the other hand, at
$t \gg 1$, the solitons running through the bonds $B_2$ and  $B_3$
are exclusively responsible for the norm. Therefore we can
naturally define transmission probabilities at $t\to +\infty$.

\begin{figure}[htb]
\centerline{\includegraphics[width=\columnwidth]{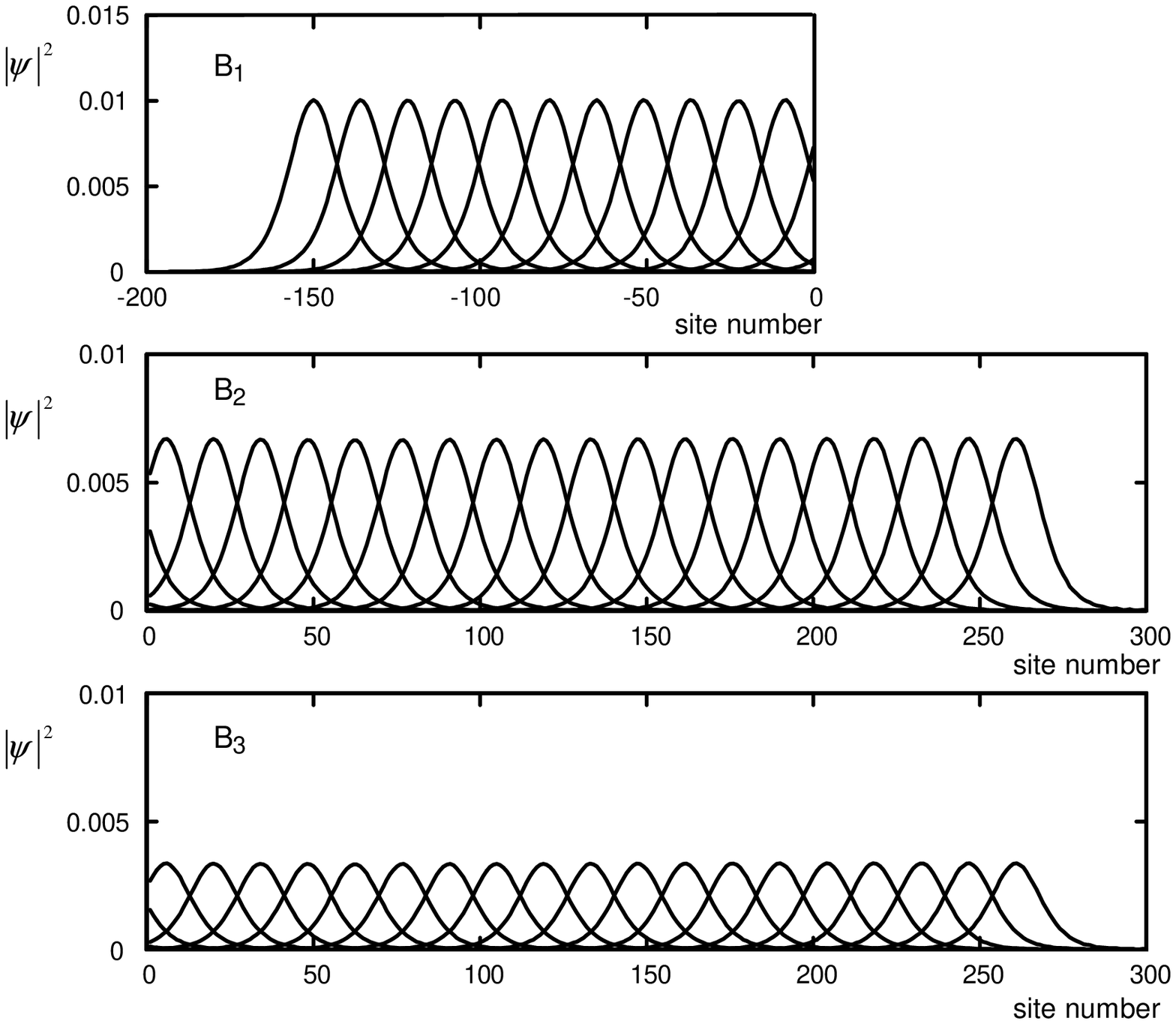}}
\centerline{\includegraphics[width=5cm,clip]{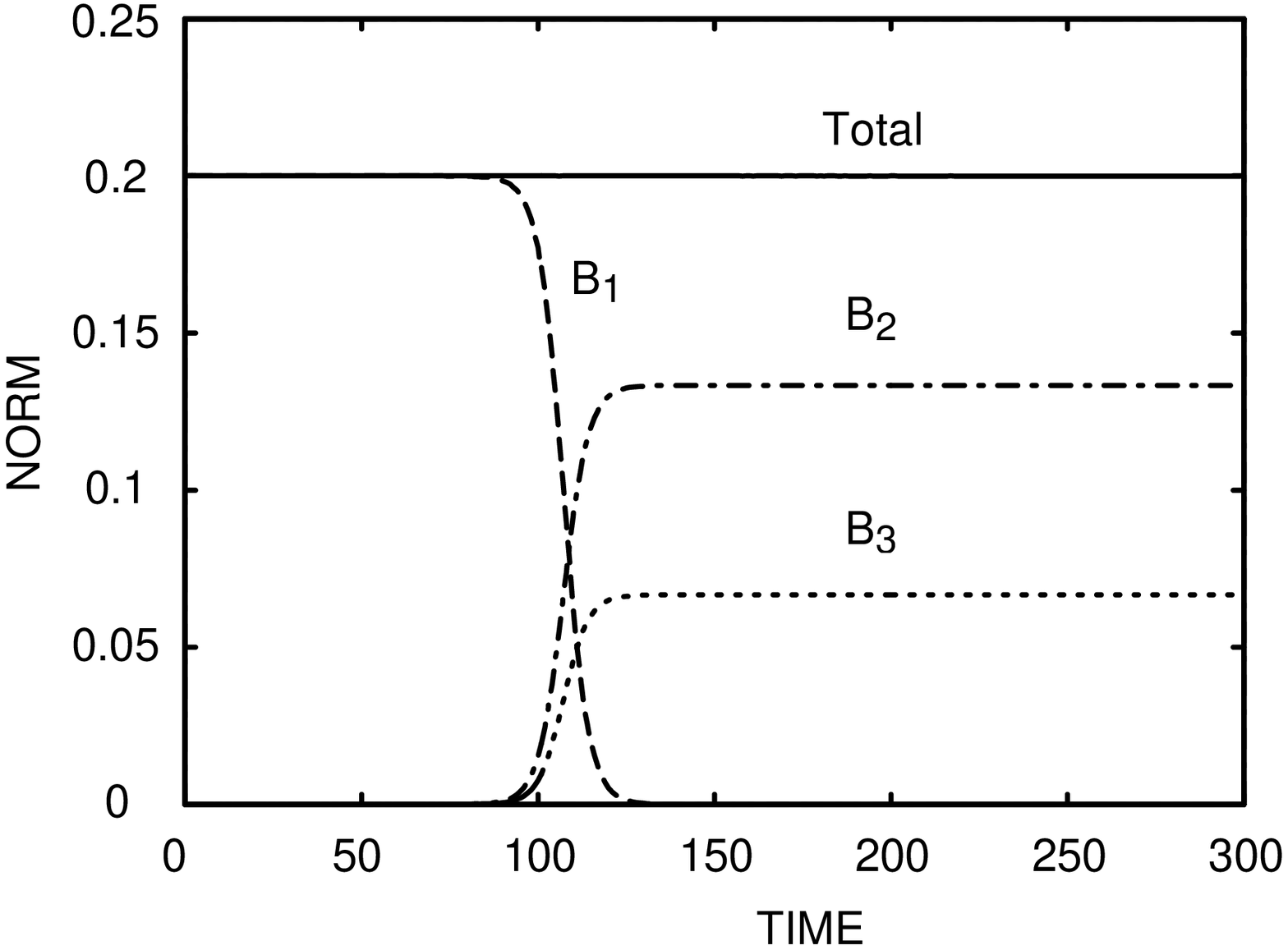}}
\caption{Numerical result for time evolution of a soliton propagation through a vertex in PSG. Strength of nonlinearity at each bond are $\gamma_{1}=1,\ \gamma_{2}=1.5,\ \gamma_{3}=3$ satisfying the sum rule in Eq.(\ref{sumrule-g}). Space distribution of wave function probability is depicted in every time interval $T=10.0$ with time used commonly in branches 2 and 3. Abscissa represents discrete lattice coordinates defined in Fig.\ref{star}. Initial profile is Ablowitz-Ladik soliton in Eq.(\ref{Sol}) at $t=0$ with parameters $\beta=0.1, \alpha=5\pi/4$. Time difference in numerical iteration is $\Delta t=0.01$ Bottom panel shows the time dependence of partial norms at each of 3 branches.
} \label{numerical}
\end{figure}

\begin{figure}[htb]
\centerline{\includegraphics[width=\columnwidth]{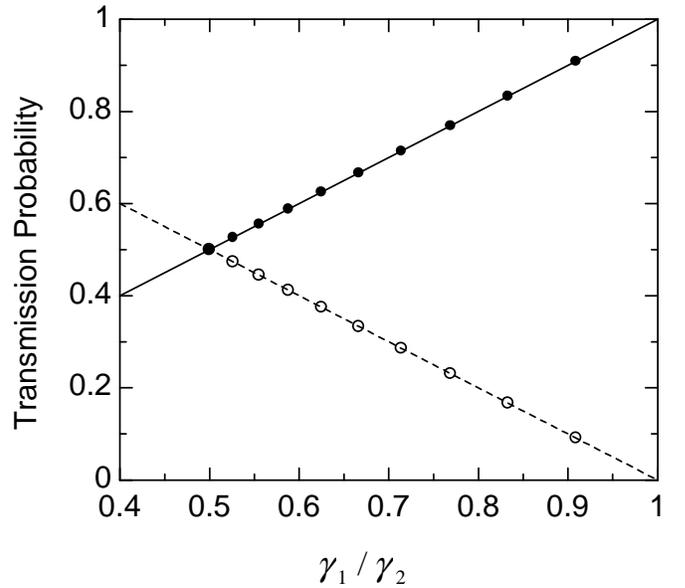}}
\caption{Transmission probabilities (TP) as a function of
$\frac{\gamma_1}{\gamma_2}$ in PSG. Symbols and lines denote numerical and theoretical
results, respectively.  A solid line with $\bullet$ and a broken
line with $\circ$ correspond to $T_2$ and $T_3$, respectively.} \label{transm}
\end{figure}

In general networks, transmission probability for an arbitrary sem-infinite bond $B_{l} (\l \ne 1)$
at discrete time $\hat{t}$ that makes $v\hat{t}$ integers are defined as
\begin{align}\label{Dtp_def}
T_{ l}&=\frac{1}{N
\gamma_{l}}\sum_{n=1}^{+\infty}\ln\left(1+\gamma_{l}|\psi_{l,n}|^2\right)=\nonumber\\
&=\frac{1}{N
\gamma_{l}}\sum_{n=1}^{+\infty}\ln\left(1+\sinh^2\beta{\rm
sech}^2(\beta(n-n_0-v\hat{t}))\right)\nonumber\\
&=\frac{\gamma_1}{N
\gamma_{l}}\sum_{n^{\prime}=1-n_0-v\hat{t}}^{+\infty}\frac{1}{\gamma_1}\ln\left(1+
\sinh^2\beta{\rm sech}^2(\beta n^{\prime})\right).
\end{align}

At $v\hat{t}\to +\infty$, $\sum_{n^{\prime}=1-n_0-v\hat{t}}^{+\infty}$ 
on the last line in Eq.(\ref{Dtp_def}) tends to
$\sum_{n^{\prime}=-\infty}^{+\infty}$ and this summation gives $N$, i.e., the normalization of the soliton in the ideal
1-d chain with the nonlinearity coefficient $\gamma_1$. Therefore
\begin{align}\label{dtp2}
T_{ l}=\frac{\gamma_1}{\gamma_{l}}
\end{align}

Under the sum rules as in Eqs. (\ref{sumrule-g}), (\ref{beta-g}) and (\ref{sumrule}) we
have the unitarity condition

\begin{align}\label{tp4}
\sum_{l=2}^{N} T_{ l}=1,
\end{align}
where the summation is taken over the semi-infinite bonds except for $B_1$. The result in Eq.(\ref{dtp2}) means that the transmission probability is inversely proportional to the strength of nonlinearity in outgoing semi-infinite bonds. 

We have checked this result using a numerical simulation of the discrete nonlinear Schr\"odinger equation(DNLSE)  on  PSG in  Fig.\ref{star}:
We numerically iterated Eqs.(\ref{eq1}), (\ref{v10}) and (\ref{vk1}) with use of Eq.(\ref{S-k}) and chose the initial profile in Eq.(\ref{Sol}) with $\gamma_1$ and $n_0=-150$ as an incoming soliton. Figure \ref{numerical} shows the result in the case that the sum rule  in Eq.(\ref{sumrule-g}) is satisfied: the soliton starting at lattice point $n=-150$ in the branch 1 enters the vertex  at $n=0$ and is smoothly split into a pair of smaller solitons in the branches 2 and 3 with no reflection at the vertex. The velocity and width of the soliton have the definite value common to all bonds, and the squared peak value of the soliton is proportional to $\gamma_k$, which are consistent with the result in Eq.(\ref{Sol}). Bottom panel in Fig.\ref{numerical} shows the time dependence of partial norms at each of 3 branches. With increasing time, the partial norms at branches 2 and 3 converge to the transmission probabilities in Eq.(\ref{dtp2}).

In Fig.\ref{transm}  transmission probabilities $T_2$ and $T_3$ are plotted as a function of $\frac{\gamma_1}{\gamma_2}$ in the wider range of $\gamma_1$ and $\gamma_1$  in the case satisfying the sum rule in Eq.(\ref{sumrule-g}).  We can confirm the linear law predicted in Eq.(\ref{dtp2}).

\begin{figure}[htb]
\centerline{\includegraphics[width=\columnwidth]{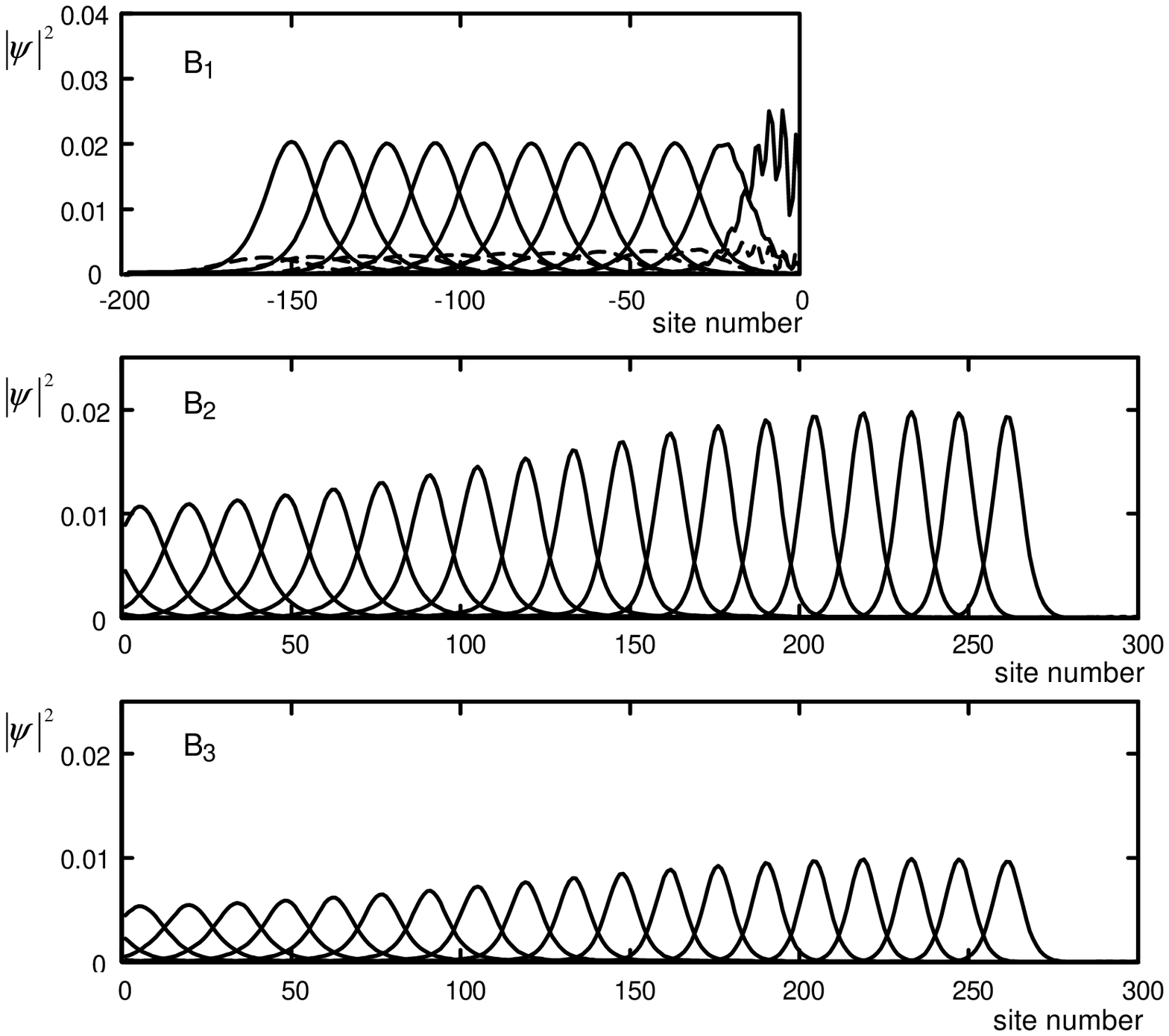}}
\centerline{\includegraphics[width=5cm,clip]{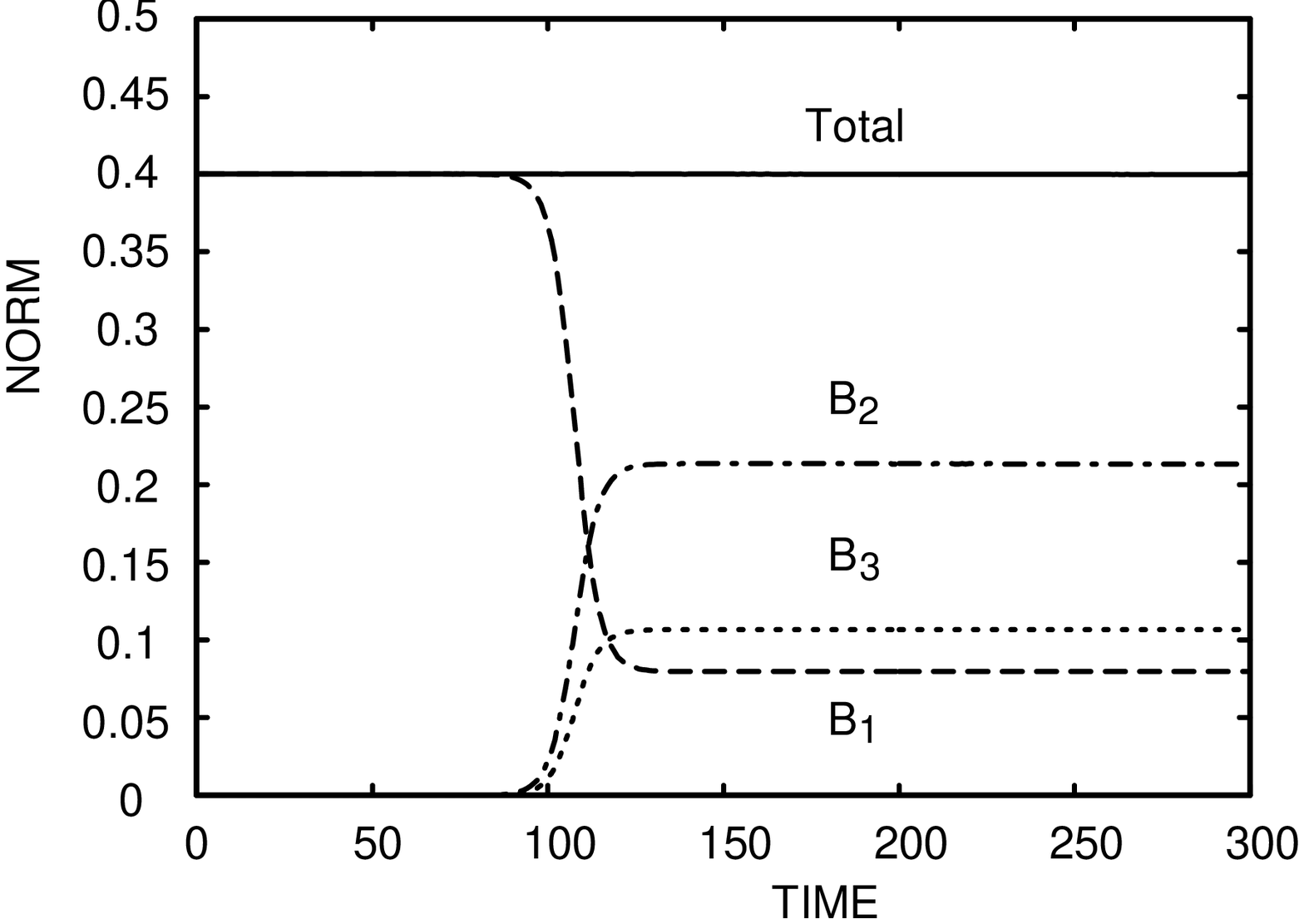}}
\caption{Numerical result for time evolution of a soliton propagation through a vertex in the case of
$\gamma_1=0.5,\gamma_2=1.5, \gamma_3=3$, which breaks  the sum rule. 
Initial profile and parameter values are the same as in Fig.\ref{numerical}. In top panel, broken curves indicate a propagation of the reflected soliton.
Bottom panel shows the time dependence of partial norms at each of 3 branches.} \label{numerical-ex}
\end{figure}

Figure \ref{numerical-ex} shows the result in the case that the sum rule  is broken: 
$\frac{\gamma_1}{\gamma_2}+\frac{\gamma_1}{\gamma_3} \ne1$. In this case the soliton starting at lattice point $n=-150$ in the branch 1 enters the vertex  at $n=0$, but is accompanied with both reflection and emergence of radiation at the vertex. It is very interesting that the velocity of the self-organized soliton have the definite value common to all bonds. In particular, the reflected soliton at the branch 1 has the same magnitude of velocity as that of the incident soliton. With increasing time, the partial norms at branches 1, 2 and 3 would converge to the reflection (on $B_1$) and transmission probabilities (on $B_2$ and $B_3$).  For some other choice of $\gamma_1$,$\gamma_2$ and $\gamma_3$ that breaks the sum rule
(which is not shown here), the asymptotically($t \gg 1$)-equal velocity of solitons running on all three semi-infinite bonds can also be observed  and provides an open question to be resolved in due course.

 
\section{Summary and discussions}
We have derived conditions under which Ablowitz-Ladik (AL) type
discrete nonlinear Schr\"odinger equation (DNLSE) on simple networks is mapped to the original one on the ideal 1-d chain and becomes completely integrable. Here the strength of cubic nonlinearity is different from bond to bond, and networks are assumed to have  at least two semi-infinite bonds with one of them used as an incoming bond. Our findings are: (1) the solution on each bond is a part of the universal (bond-independent) soliton solution of the completely-integrable DNLSE on the 1-d chain, but is multiplied by the inverse of square root of bond-dependent nonlinearity; (2) the inverse nonlinearity at an incoming bond should be equal to the sum of inverse nonlinearities at the remaining outgoing bonds; (3) with use of the above two findings, there exist an infinite number of constants of motion. 
The parameters $s_2$ and $s_3$,  which played an essential role in deriving the connection formula, 
are introduced to define the inter-site interaction at the vertex and are not obtained  from the norm and energy conservations, in marked contrast to the case of networks consisting of continuum segments\cite{sob}.
The argument on a branched chain or a primary star graph (PSG) is generalized to general star graphs and  tree graphs by using the induction method. As a practical issue, with use of AL soliton injected  
through the incoming bond, we obtain transmission probabilities inversely proportional to the strength of nonlinearity on the outgoing bonds.

{\em Acknowledgments.} K.N. is grateful to F. Abdullaev for suggesting a significance of extending our preceding work on the continuum NLSE to the discrete case. 



\end{document}